\documentclass[lettersize,journal]{IEEEtran}
\usepackage{amsmath,amsfonts}
\usepackage{algorithmic}
\usepackage{algorithm}
\usepackage{array}
\usepackage[caption=false,font=small,labelfont=rm,textfont=sf]{subfig}
\usepackage{textcomp}
\usepackage{stfloats}
\usepackage{url}
\usepackage{verbatim}
\usepackage{graphicx}
\usepackage{cite,color}
\usepackage{booktabs}
\usepackage{multirow}
\graphicspath{{figure/}} 
\begin{document}

\title{\LARGE
Transformer-based Hybrid Beamforming with Dynamic Subarray for Near-Space Airship-Borne Communications}

\author{Ruiqi Wang, Zhen Gao, \IEEEmembership{Member,~IEEE}, Keke Ying, Ziwei Wan, Symeon Chatzinotas, \IEEEmembership{Fellow,~IEEE}, and Mohamed-Slim Alouini, \IEEEmembership{Fellow,~IEEE}
\thanks{The work was supported in part by the Natural Science Foundation of China under Grant 62471036, in part by Shandong Province Natural Science Foundation under Grant ZR2025QA30, in part by Beijing Natural Science Foundation under Grants L242011.  \textit{(Corresponding authors: Zhen Gao and Ziwei Wan.)}
\par
Ruiqi Wang and Keke Ying are with the School of Information and Electronics, Beijing Institute of Technology, Beijing 100081, China, also with the State Key Laboratory of Environment Characteristics and Effects for Near-space, Beijing 100081, China (e-mails: rachem@126.com, ykk@bit.edu.cn).
\par
Zhen Gao is with  Beijing Institute of Technology (BIT), Zhuhai 519088, China, also with the State Key Laboratory of CNS/ATM, Beijing 100081, China, also with the Advanced Technology Research Institute, BIT, Jinan 250307, China  (e-mail: gaozhen16@bit.edu.cn).
\par
Z. Wan is with the Yangtze Delta Region Academy of Beijing Institute of Technology (Jiaxing), Jiaxing 31401 (e-mail: ziweiwan@bit.edu.cn).
\par
Symeon Chatzinotas is with the Interdisciplinary Center for Security, Reliability and Trust (SnT) - University of Luxembourg, L-1855 Luxembourg (e-mail: Symeon.Chatzinotas@uni.lu).
\par
M-S. Alouini is with the Computer, Electrical and Math ematical Sciences and Engineering Division, King Abdullah University of
Science and Technology, Thuwal, 23955, Kingdom of Saudi Arabia (e-mail: slim.alouini@kaust.edu.sa).
}
\vspace{-7mm}
}



\maketitle

\begin{abstract}
This paper proposes a hybrid beamforming framework for massive multiple-input multiple-output (MIMO) in near-space airship-borne communications. To achieve high energy efficiency (EE) in energy-constraint airships, a dynamic subarray structure is introduced, where each radio frequency chain (RFC) is connected to a disjoint subset of the antennas according to channel state information (CSI). The proposed joint dynamic hybrid beamforming network (DyHBFNet) comprises three key components: 1) An analog beamforming network (ABFNet) that optimizes the analog beamforming matrices and provides auxiliary information for the antenna selection network (ASNet) design, 2) an ASNet that dynamically optimizes the connections between antennas and RFCs, and 3) a digital beamforming network (DBFNet) that optimizes digital beamforming matrices by employing a model-driven weighted minimum mean square error algorithm for improving beamforming performance and convergence speed. The proposed ABFNet, ASNet, and DBFNet are all designed based on advanced Transformer encoders. Simulation results demonstrate that the proposed framework significantly enhances spectral efficiency and EE compared to baseline schemes. Additionally, its robust performance under imperfect CSI makes it a scalable solution for practical implementations.
\end{abstract}

\vspace{-3mm}
\section{Introduction}
Near-space airship-borne communications stand out as a promising technology in sixth-generation (6G), since near-space airships can realize extended coverage while providing lower latency due to lower altitudes (20-100 km) compared to popular low-Earth-orbit satellites \cite{SAGSIN}. To compensate for the severe path loss in long-range air-to-ground links, massive multiple-input multiple-output (MIMO) offering substantial array gain is an indispensable technique for airship-borne communications \cite{MIMO}. Moreover, near-space airships are usually powered by solar panels and thus not charged frequently, which causes energy shortage for the communication tasks. This motivates the use of hybrid analog-digital (HAD) architecture \cite{HBF} for near-space airship massive MIMO.\par
It has been reported that HAD massive MIMO with dynamic subarray structure can strike a balance between the system performance and hardware complexity\cite{Dynamic Subarrays}. In such an array architecture, the connection between the antennas and radio frequency chains (RFCs) can be adjusted according to the channel state information (CSI). To resolve the hybrid beamforming (HBF) problem under dynamic structure, an iterative algorithm based on the block coordinate descent method was proposed in \cite{Dynamic hybrid beamforming} for single-user scenario. Furthermore, the multi-user scenario was considered in \cite{multi-user}, where a two-stage HBF design using Butler matrices and weighted minimum mean square error (WMMSE) algorithm was proposed. However, the aforementioned methods have high computational complexity and poor adaptability, and their performance will decline significantly under imperfect CSI\cite{Transformer}. As a remedy, deep learning (DL)-based methods has been widely applied in the field of HBF. For example, the authors of \cite{CNN} proposed a deep learning approach for joint antenna selection and HBF under dynamic structure. {To avoid labeling a large amount of data during training stage, deep unsupervised learning was introduced in \cite{Deep Unsupervised Learning} for HBF design. However, the purely data-driven approaches fail to effectively integrate domain knowledge of conventional beamforming algorithm.} Therefore, the model-driven network is a better choice for HBF. Additionally, the Transformer architecture utilizes self-attention layers integrated with a refined multi-head mechanism, demonstrating superior performance over other network models in various scenarios \cite{attention}. To the best of our knowledge, the interplay between the Transformer and the HBF under dynamic subarray structure has not been well studied.
\par In this paper, we propose a Transformer-based dynamic hybrid beamforming network (DyHBFNet) for HAD architecture with dynamic subarray. The proposed DyHBFNet comprises three key components: 
\vspace{-0.3mm}
\begin{itemize}
    \item An analog beamforming network (ABFNet) that optimizes the analog beamforming matrices and provides auxiliary information for the subsequent antenna selection network (ASNet).
    \item An ASNet that utilizes the analog beamforming information provided by ABFNet to dynamically optimize the connections between antenna elements and RFCs.
    \item A digital beamforming network (DBFNet) that optimizes digital beamforming matrices by employing a model-driven WMMSE algorithm for improving beamforming and convergence performance.
\end{itemize}
Moreover, simulation and comparative study show that the proposed scheme exhibits remarkable resilience under imperfect CSI, which makes it an excellent candidate for the beamforming design in airship-borne communications.
\par \textit{Notations}: This paper uses lower-case letters for scalars, lower-case bold letters for column vectors, and upper-case bold letters for matrices. Superscripts $(\cdot)^*$, $(\cdot)^T$, $(\cdot)^H$, $(\cdot)^{-1}$ denote the conjugate, transpose, conjugate transpose and inversion operators, respectively. $\| \mathbf{A} \|_F$ is the Frobenius norm of $\mathbf{A}$ and $\| \mathbf{a} \|_0$ is the $l_0$ norm of $\mathbf{a}$. $\text{vec}(\mathbf{A})$ denotes the vectorization operation. $\Re\{\cdot\}$ and $\Im\{\cdot\}$ denote the real part and imaginary part of the corresponding arguments, respectively. The $(i, j)$-th entry of $\mathbf{A}$ is $[\mathbf{A}]_{i,j}$, and $[\mathbf{A}]_{i,:}$($[\mathbf{A}]_{:,j}$ ) denotes the $i$-th row ($j$-th column) of $\mathbf{A}$. $[\mathbf{A}]_{:,i:j}$ is the sub-matrix containing the $i$-th to $j$-th columns of $\mathbf{A}$. The mathematical expectation is denoted by $\mathbb{E}(\cdot)$. $\mathbf{A}\odot\mathbf{B}$ denotes the Hadamard product of $\mathbf{A}$ and $\mathbf{B}$.
\vspace{-3mm}
\section{System Model}
\begin{figure}[!t]
\centering
\includegraphics[width=0.2\textwidth]{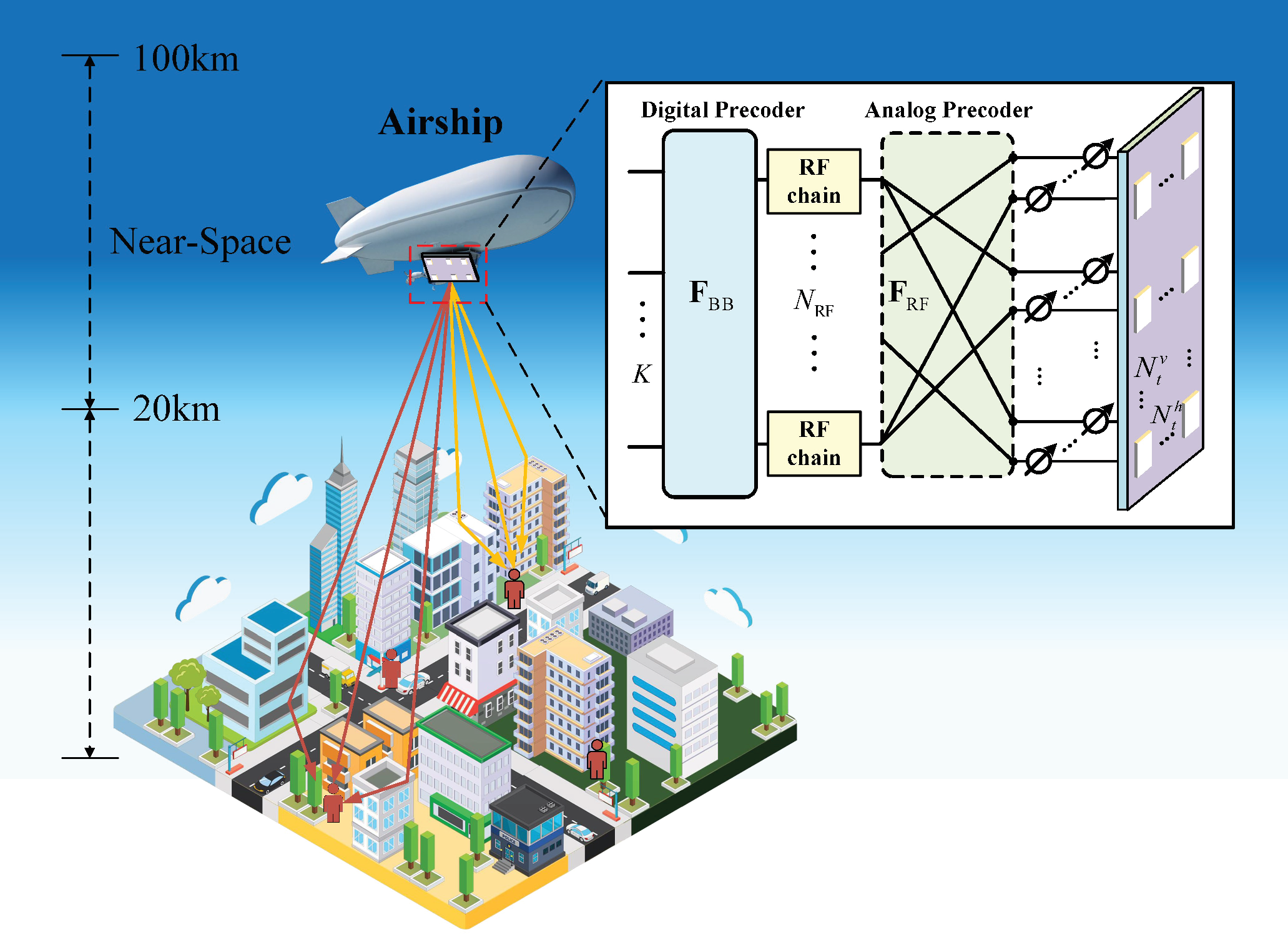}
\vspace{-3mm}
\caption{Near-space airship-borne communication system assisted by HAD massive MIMO with dynamic subarray structure.}
\hfil
\label{stratosat}
\vspace{-4mm}
\end{figure}
We consider a downlink multi-user communication system assisted by near-space airships with HAD massive MIMO. As illustrated in Fig.~\ref{stratosat}, one aerial BS deployed at a near-space airship transmits $K$ data streams to $K$ single-antenna terrestrial users. The aerial BS utilizes a uniform planar array (UPA) with $N_t=N_t^v \times N_t^h$ antennas and $N_\text{RF}$ RFCs, where $N_\text{RF}\ll N_t$, and $N_t^v$ and $N_t^h$ are the numbers of antennas along vertical and horizontal directions, respectively. For simplicity, we assume $N_\text{RF} = K$. To cope with the energy-constraint property of near-space airships, we introduce the dynamic subarray structure, i.e., each antenna is exclusively connected to a single RFC via a phase shifter (PS). We consider orthogonal frequency division multiplexing (OFDM) with $N_c$ subcarriers. The signal received by the $k$-th user ($1\leq k\leq K$) associated with the $m$-th subcarrier ($1\leq m\leq N_c$) can be expressed as
\begin{equation}
y[k, m] = \mathbf{h}^T[k, m] \mathbf{F}_\text{RF}\mathbf{F}_{\text{BB}}[m] \mathbf{s}[m] + n[k, m],
\label{recived_signal}
\end{equation}
where $\mathbf{h}[k, m] \in \mathbb{C}^{N_t\times 1}$ is the channel vector between the BS and the $k$-th user equipment (UE) associated with the $m$-th subcarrier, ${\mathbf{F}}_{\text{RF}} \in \mathbb{C}^{N_t\times K} $ is the analog beamforming matrix of the proposed dynamic subarray structure, $\mathbf{F}_{\text{BB}}[m]\in \mathbb{C}^{K\times K}$ is the baseband digital beamforming matrix, $\mathbf{s}[m] \in \mathbb{C}^{K \times 1}$ is the transmitted signal vector, which satisfies $\mathbb{E}(\mathbf{s}[m]\mathbf{s}[m]^H) = \mathbf{I}_K$, and $n[k, m]\sim \mathcal{CN}(0,\sigma^2_n)$ is additive white Gaussian noise (AWGN). Under the considered dynamic subarray structure, the analog beamforming matrix can be expressed as $\mathbf{F}_\text{RF} = \tilde{\mathbf{F}}_{\text{RF}} \odot \mathbf{X}_{\text{sel}}$, where $\tilde{\mathbf{F}}_{\text{RF}}$ is the analog beamforming matrix of fully-connected structure, $\mathbf{X}_{\text{sel}} \in \{0,1\}^{N_t\times K} $ is the antenna selection matrix and each row of the matrix satisfies $ \Vert [\mathbf{X}_{\text{sel}}]_{i,:}\Vert_0 = 1$. It is worth noting that the fully-connected structure analog beamforming matrix should satisfy the unit modulus constraint, i.e., $\left|[\tilde{\mathbf{F}}_\text{RF}]_{i,j} \right|=1/\sqrt{N_t}$. Moreover, the digital beamforming matrix should satisfy the transmit power constraint, i.e., $\Vert \mathbf{F}_\text{BB}[m]\Vert_{F}^2 = P_t$, where $P_t$ is the transmit power allocated to each subcarrier.\par

\begin{figure*}[b!]
	\vspace*{-3mm}
	\noindent\rule[0.25\baselineskip]{\textwidth}{1pt}
\begin{equation}
\centering
\begin{aligned}
\mathbf{h}[k, m]=\sqrt{F_k}\Bigg[\mathbf{a}_t(\theta^{k}_\text{LoS}, \phi^{k}_\text{LoS}) +\sqrt{\frac{1}{L_p}}\sum_{l=1}^{L_p} \alpha_{l,k} \mathbf{a}_t(\theta^{l,k}_\text{NLoS}, \phi^{l,k}_\text{NLoS}) e^{-j \frac{2\pi m \tau^{l,k}_\text{NLoS}}{N_c T_s}} \Bigg],
\end{aligned}
\label{channel}
\end{equation}
\end{figure*}

\vspace{-3mm}
\section{Proposed Joint DyHBFNet}    
\begin{figure}[!t]
\centering
\includegraphics[width=0.3\textwidth]{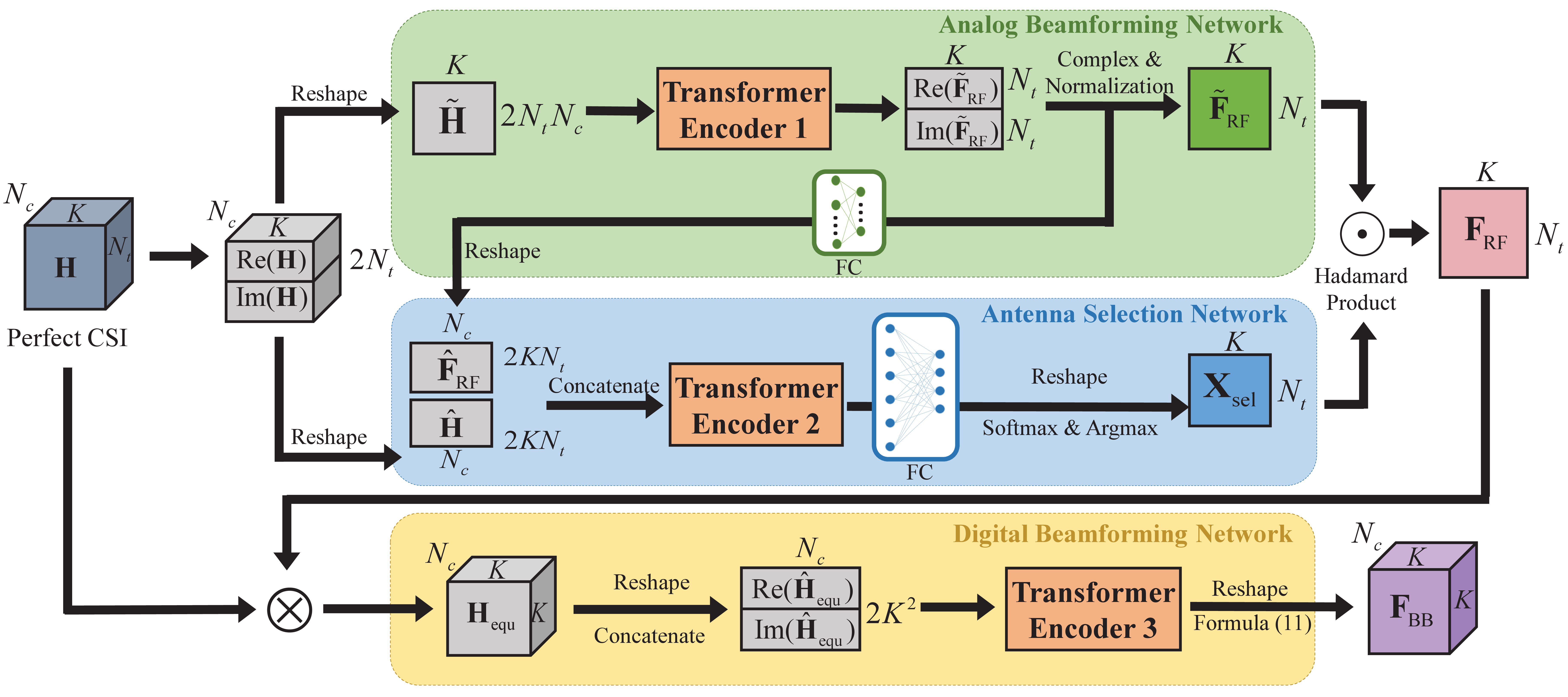}
\vspace{-4mm}
\caption{Structure of the proposed joint DyHBFNet.}
\hfil
\label{net}
\vspace{-7mm}
\end{figure}

In this section, we introduce the Transformer-based joint DyHBFNet. First, we characterize the channel model and formulate the corresponding optimization objective. Then, we propose the ABFNet for analog beamforming and providing auxiliary information for ASNet. Next, we develop the ASNet to design the connections between antennas and RFCs by integrating analog beamforming information to meet the requirement of dynamic subarray structure. In the end, for the DBFNet, we employ a model-driven WMMSE algorithm to improve beamforming performance and reduce computational complexity. The proposed DyHBFNet are jointly trained to maximize SE in an end-to-end manner. The block diagram of the proposed DyHBFNet is illustrated in Fig.~\ref{net}.
\vspace{-3mm}
\subsection{Problem Formulation}
Considering that the BS is deployed at a high altitude, compared with terrestrial base station communication, the channel $\mathbf{h}[k, m]$ consists of a line-of-sight (LoS) path and several non-line-of-sight (NLoS) paths, where the angles of departure (AoDs) of the NLoS paths display a small angular spread around the AoD of the LoS path. Assuming that the number of multipath components is $L_p$, the channel $\mathbf{h}[k, m]$ between the BS and the $k$-th UE associated with the $m$-th subcarrier can be expressed as shown in \eqref{channel}, where $F_k= \left( \frac{\lambda}{4\pi r_k}\right)^2$ represents the path loss with respect to large-scale fading, $\lambda$ is the wavelength, $r_k$ is the distance between the BS and the $k$-th UE, and $\alpha_{l,k}\sim \mathcal{CN}(0,1)$ represents the small-scale gain of the $l$-th path. The azimuth and elevation angles of departure (AoDs) for the LoS path are denoted by $\theta^{k}_\text{LoS}$ and $\phi^{k}_\text{LoS}$, respectively, while $\theta^{l,k}_\text{NLoS}$ and $\phi^{l,k}_\text{NLoS}$ are the azimuth and elevation AoDs of the $l$-th NLoS path, {which are distributed around $\theta^{k}_\text{LoS}$ and $\phi^{k}_\text{LoS}$ with small angular spread.} The relative delay of the $l$-th NLoS path is denoted by $\tau^{l,k}_\text{NLoS}$. Furthermore, $\mathbf{a}_t(\cdot)$ denotes the normalized transmit array response vector, which can be modeled as
\begin{equation}
\label{array_vector}
\begin{aligned}
\mathbf{a}_t(\theta, \phi)  =& [ 1, \dots, e^{j\frac{2\pi}{\lambda}d(m_1 \sin\theta\cos\phi + m_2 \sin\phi)}, \\
&\dots, e^{j\frac{2\pi}{\lambda}d((N_t^v - 1)\sin\theta\cos\phi + (N_t^h - 1)\sin\phi)} ],
\end{aligned}
\end{equation}
where $0\le m_1\le N_t^v-1$, $0\le m_2\le N_t^h-1$, and $d = \lambda/2$ is the adjacent antenna spacing.\par
A common optimization objective for HBF problem in airship-borne communication system is maximizing spectral efficiency (SE), which can be summarized as
\begin{align}
\max_{\mathbf{F}_{\text{RF}},\mathbf{F}_{\text{BB}}[m]} \quad & R = \frac{1}{N_c} \sum_{k=1}^{K} \sum_{m=1}^{N_c} \log_{2}(1+\text{SINR}[k,m]) \nonumber \\
\text{s.t.} \quad 
&\left|[\tilde{ \mathbf{F} }_{\text{RF}}]_{i,j} \right|=\frac{1}{\sqrt{N_t}}, \forall i, j, \nonumber \\
&\Vert [\mathbf{X}_{\text{sel}}]_{i,:}\Vert_0 = 1,\forall i, \ \| \mathbf{F}_{\text{BB}}[m] \|_F^2 = P_t, \, \forall m.
\label{optimization}
\end{align}
The signal-to-interference-plus-noise ratio (SINR) of the $k$-th UE associated with the $m$-th subcarrier can be expressed as
\begin{equation}
\label{SINR}
\text{SINR}[k,m] = 
\frac{\left| \mathbf{h}^T[k,m] \mathbf{F}_\text{RF}  \mathbf{f}_\text{BB}[k,m] \right|^2}
{\sum\limits_{i=1, i \neq k}^K \left| \mathbf{h}^T[k,m]  \mathbf{F}_\text{RF} \mathbf{f}_\text{BB}[i,m] \right|^2 + \sigma_m^2},
\end{equation}
where $\mathbf{f}_\text{BB}[k,m] \in \mathbb{C}^{K\times 1}$ is the $k$-th column of $\mathbf{F}_{\text{BB}}[m]$. 

\vspace{-3mm}
\subsection{Proposed ABFNet for Analog Beamforming }
For the ABFNet, we first convert the 3-D complex-valued CSI $\mathbf{H} \in \mathbb{C}^{N_c \times K \times N_t}$ into a real-valued matrix by concatenating  $\Re\{{\mathbf{H}}\}$ and $\Im{\{\mathbf{H}\}}$. This matrix is subsequently reshaped to get $\tilde{\mathbf{H}} \in \mathbb{R}^{K\times 2N_tN_c}$. Specifically, for the $k$-th ($k = 1, 2, \ldots, K$) row of $\tilde{\mathbf{H}}$, its first $N_tN_c$ entries, i.e. $\left[\tilde{\mathbf{H}} \right]_{k,1:N_tN_c}$, are $\operatorname{vec}( \Re\left\{\mathbf{H}[k] \right\})$, while the next $N_tN_c$ entries are $\operatorname{vec}(\Im\left\{\mathbf{H}[k] \right\})$.  $\tilde{\mathbf{H}}$ serves as the input to Transformer Encoder 1. Here, $K$ denotes the input sequence length for the Transformer to extract the correlation between the RFCs.\par
Within the Transformer, the input sequence is first mapped to a sequence of vectors with a dimension of 256 using a fully connected (FC) linear layer followed by a positional embedding layer. The positional embedding employs sine functions at different frequencies to encode the positions of different subcarriers. The Transformer then processes this sequence through three identical layers, each consisting of a multi-head self-attention sublayer and a multilayer perception sublayer to extract high-level features from the input sequence.\par
Then Transformer Encoder 1 generates the real and imaginary parts of $\tilde{\mathbf{F}}_{\text{RF}}$. By applying a complex function and performing amplitude normalization, the analog beamforming matrix that satisfies the unit modulus constraint is obtained. Meanwhile, the $\Re{\{\tilde{\mathbf{F}}_\text{RF}}\}$ and $\Im{\{\tilde{\mathbf{F}}_\text{RF}}\}$ are processed by an FC layer and then reshaped into $\hat{\mathbf{F}}_\text{RF} \in \mathbb{R}^{N_c\times 2KN_t}$ to provide analog beamforming information for the ASNet design. 
\vspace{-3mm}
\subsection{Proposed ASNet for Antenna Selection }
The ASNet is designed to select the optimal connections between antennas and RFCs, aiming to maximizing SE under the hardware constraint of dynamic subarray structure. Given $\Vert [\mathbf{X}{_\text{sel}}]_{i,:}\Vert_0 = 1$, the design of $\mathbf{X}_\text{sel}$ can be formulated as $N_t$ multi-class classification tasks, where each of the $N_t$ antennas is assigned to a single RFC. First, the real-valued perfect CSI is reshaped into a matrix $\hat{\mathbf{H}} \in \mathbb{R}^{N_c \times 2KN_t}$. Then, $\hat{\mathbf{H}}$ is concatenated with $\hat{\mathbf{F}}_\text{RF}$ output from Transformer Encoder 1, forming a real-valued sequence of size $N_c \times 4KN_t$ as the input to Transformer Encoder 2. Here, $N_c$ denotes the effective input sequence length. The extracted features with dimension of $N_c\times KN_t $ are then passed through a FC linear layer and reshaped to obtain the matrix $\tilde{\mathbf{X}}_\text{sel} \in \mathbb{R}^{N_t \times K}$. Subsequently, the softmax activation function is applied to convert $\tilde{\mathbf{X}}_\text{sel}$ into a probability matrix. To get the antenna selection matrix, we use the argmax function to determine the maximum value in each row of the matrix $\mathbf{X}_\text{sel}$. More particularly, for the $(i,j)$-th entry of $\mathbf{X}_\text{sel}$, it equals 1 only when $j = \mathop{\arg\max}\limits_{j}(\text{softmax}([\tilde{\mathbf{X}}_\text{sel}]_{i,:}))$ and remains 0 in other cases. Finally, $\mathbf{F}_{\text{RF}}$ is obtained by the Hadamard product of $\tilde{\mathbf{F}}_{\text{RF}}$ and $\mathbf{X}_\text{sel}$. 

\vspace{-3mm}
\subsection{Proposed DBFNet for Digital Beamforming}
For the digital beamforming, we employ the low-complexity model-driven WMMSE algorithm to achieve improved beamforming performance. With the analog beamforming matrix $\mathbf{F}_{\text{RF}}$ designed by ABFNet and ASNet, and given the CSI $\mathbf{h}[k,m]$, the BS can obtain the equivalent baseband CSI matrix as $\mathbf{h}^T_\text{equ}[k,m] =\mathbf{h}^T[k,m]\mathbf{F}_{\text{RF}}$. Then the optimization problem in (\ref{optimization}) can be simplified as 
\begin{align}
\max_{\mathbf{F}_{\text{BB}}[m]} \quad & R = \frac{1}{N_c} \sum_{k=1}^{K} \sum_{m=1}^{N_c} R[k,m] \nonumber \\
\text{s.t.} \quad 
& \| \mathbf{F}_{\text{BB}}[m] \|_F^2 = P_t, \, \forall m.
\label{sum-rate maximization}
\end{align}
In traditional WMMSE algorithm, it has been proved that the SE maximization problem in (\ref{sum-rate maximization}) is equivalent to a sum-mean square error minimization problem\cite{WMMSE}, and thus problem (\ref{sum-rate maximization}) can be further separated into three convex sub-problems which are solved alternatively to obtain a local optima of digital beamforming matrix. Specifically, the alternative optimization of WMMSE can be summarized as follows
\begin{align}
u[k,m]&=\frac{\mathbf{h}_{\mathrm{equ}}^T[k,m]\mathbf{f}_{\mathrm{BB}}[k,m] }{ \sum\limits_{l = 1,l\neq k}^{K}\left|{\mathbf{h}}_{\mathrm{equ}}^T[k,m]\mathbf{f}_{\mathrm{BB}}[l,m]\right|^2+\sigma_{m}^{2}},\\
w[k,m]&=(1 - \frac{\left|\mathbf{h}_{\mathrm{equ}}^T[k,m]\mathbf{f}_{\mathrm{BB}}[k,m]\right|^{2}}{ \sum\limits_{l = 1,l\neq k}^{K}\left|{\mathbf{h}}_{\mathrm{equ}}^T[k,m]\mathbf{f}_{\mathrm{BB}}[l,m]\right|^2+\sigma_{m}^{2}})^{-1},\\
\mathbf{f}_{\text{BB}}[k,m] &= \bigg(\mu[m]\mathbf{I}_{K} + \sum\limits_{l=1}^{K} w[l,m] \left|u[l,m]\right|^2{\mathbf{h}}_{\text{equ}}[l,m] \nonumber\\
&\times{\mathbf{h}}_{\text{equ}}^H[l,m] \bigg)^{-1}{u[k,m] w[k,m]{\mathbf{h}}_{\text{equ}}[k,m]}\label{f},
\end{align}
where $u[k,m]$ and $w[k,m]$ are the corresponding auxiliary factors in WMMSE algorithm. Additionally, $\mu[m]$ is a Lagrange multiplier which can be computed by algorithm in \cite{DL-WMMSE}.\par
Although the traditional WMMSE algorithm can theoretically converge to a relatively good solution, it needs a large number of iterations and may have a slow convergence speed, particularly in complex scenarios such as massive MIMO systems or situations with a large number of users \cite{DL-WMMSE}. To this end, we propose a Transformer-based model-driven WMMSE algorithm \cite{wmh} for designing $\textbf{f}_\text{BB}[k,m]$, which can greatly reduce the computational complexity. Based on (\ref{f}), the optima for $\mathbf{f}_\text{BB}[k,m]$ takes the following form
\begin{align}
    \mathbf{f}_{\text{BB}}[k,m] = &\bigg(b[m]\mathbf{I}_{K} + \sum_{l=1}^{K} c[l,m]  {\mathbf{h}}_{\text{equ}}[l,m] {\mathbf{h}}_{\text{equ}}^H[l,m] \bigg)^{-1}\nonumber\\
&\times {a[k,m] {\mathbf{h}}_{\text{equ}}[k,m]},
\label{fBB}
\end{align} 
which indicates that the optimal digital beamforming matrix depends on parameters
$\{a[k, m], b[m],c[k, m], \, \forall k, \forall m\}$. To obtain their value without bulky iterative algorithm, we propose a model-driven network. Specifically, Transformer Encoder 3 takes the equivalent channel $\mathbf{H}_\text{equ}\in \mathbb{C}^{Nc\times K \times K}$ as the input, where $\mathbf{H}_\text{equ}[m] = \mathbf{H}[m]\mathbf{F}_\text{RF},\ m=1,2,\cdots,N_c$. Then, we converts $\mathbf{H}_\text{equ}$ into a 1D real-valued input sequence by reshaping and concatenating $\Re{\{\mathbf{H}_\text{equ}\}}$ and $\Im{\{\mathbf{H}_\text{equ}\}}$. The sequence is then processed by Transformer Encoder 3 to output the key parameters in the proposed WMMSE scheme. Formula (\ref{fBB}) is then used to obtain the $\mathbf{F}_\text{BB}[m], \ m=1,2,\cdots,N_c$. Finally, we impose the power constraint on the digital beamforming matrix, i.e., $\mathbf{F}_\text{BB}[m] \leftarrow \sqrt{P_t}\frac{\mathbf{F}_{\text{BB}}[m]}{\|\mathbf{F}_{\text{BB}}[m]\|_F}$. For the proposed DyHBFNet, we choose the negative SE as the loss function, given by 
$L_\mathrm{loss} = -\frac{1}{N_c} \sum_{k=1}^{K} \sum_{m=1}^{N_c} R[k,m]$.

\section{Simulation Results}

\begin{figure}[!t]
\centering
\subfloat[]{
\includegraphics[width=0.2\textwidth]{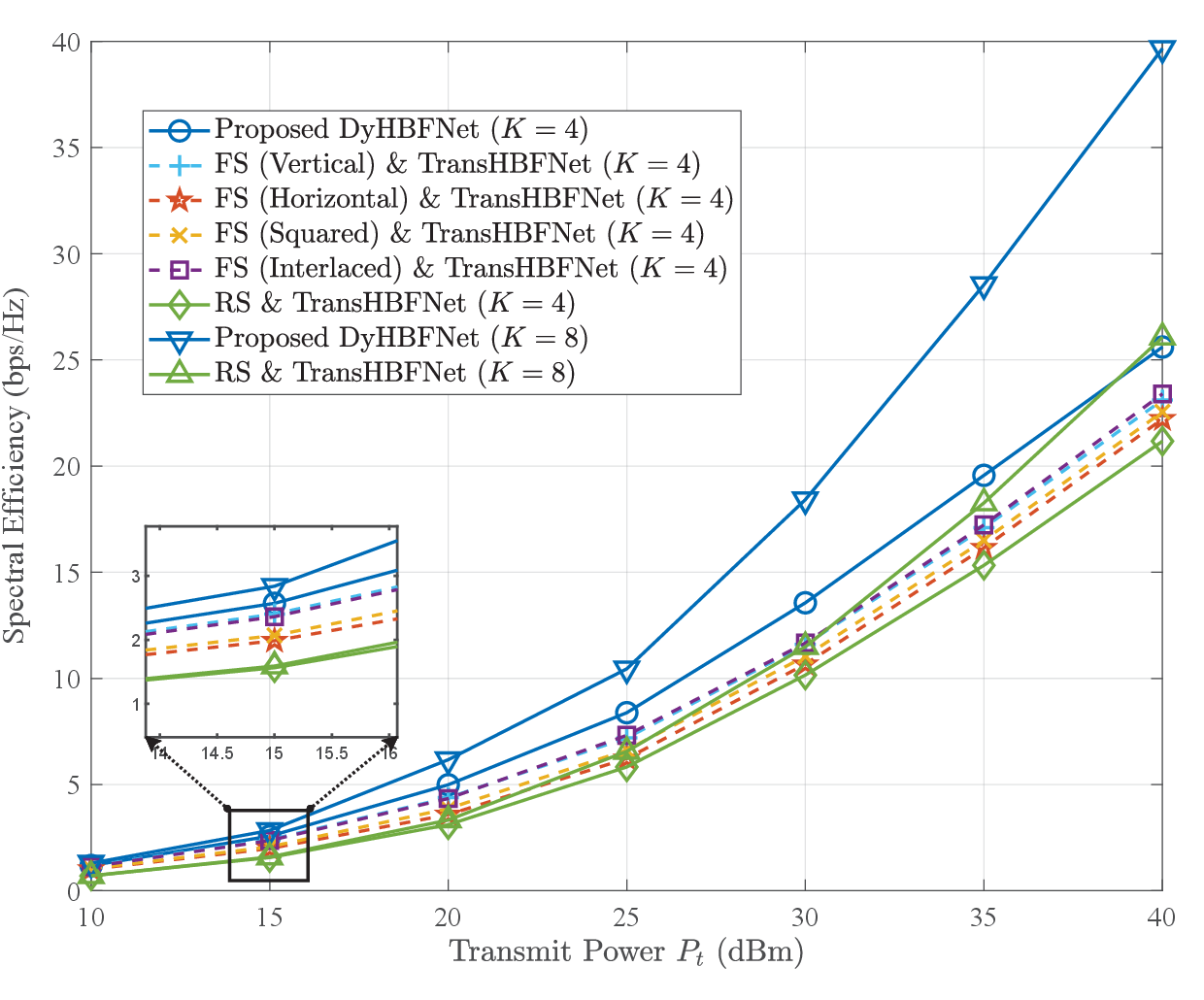}
\label{fig_first_case}
}
\vspace{-2mm}
\hfil
\subfloat[]{
\includegraphics[width=0.2\textwidth]{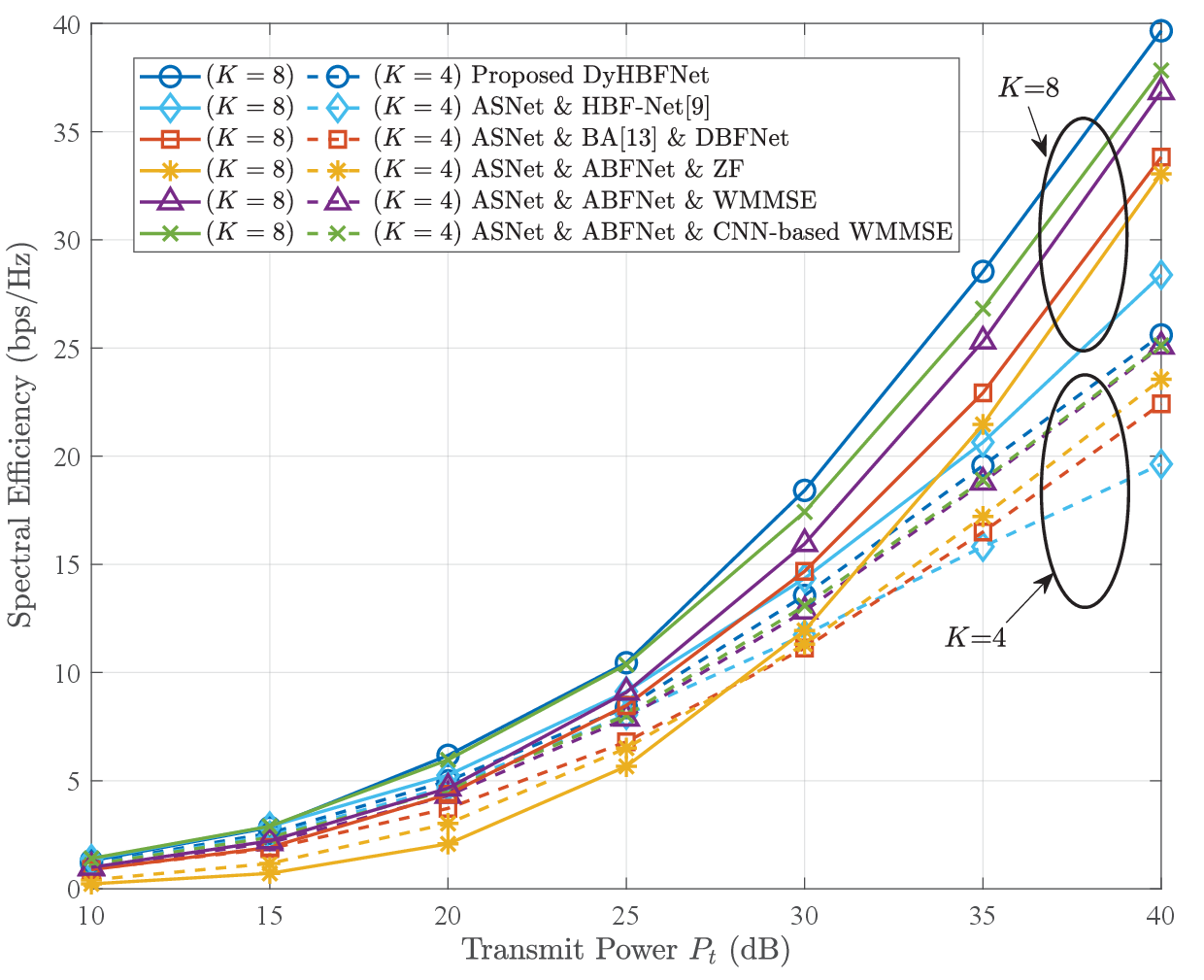}
\label{fig_second_case}
}
\caption{SE performance comparison versus $P_t$. (a) Different dynamic subarray schemes with TransHBFNet. (b) ASNet with different beamforming schemes.}
\label{SE performance}
\vspace{-4mm}
\end{figure}
In this section, we evaluate the SE and energy efficiency (EE) performance of the proposed joint DyHBFNet.
\vspace{-3mm}
\subsection{Simulation Settings}
In the simulations, we adopt the typical multipath MIMO-OFDM channel model described in Section II. The BS, equipped with an $8 \times 8$ UPA, serves $K$ single-antenna UEs. The carrier frequency is $f_c = 3 \, \text{GHz}$, the bandwidth is $B_s = 30 \, \text{MHz}$, and the noise power spectral density is $-174 \, \text{dBm/Hz}$. For the LoS path of the channel, the azimuth and elevation AoD follow a uniform distribution $\mathcal{U}[-\pi/3, \pi/3]$. For the NLoS paths, the number of paths is $L_p = 30$, the path delays are uniformly distributed in $[0, 8T_s]$, where $T_s = 1/B_s$ denotes the symbol period, and the angle spread is $\Delta\theta = \pm 10^\circ$. Besides, we set $N_c = 32$, $r_k=25\,\text{km}$. The generated channel dataset is divided into training, validation, and testing subsets, containing 204,800, 20,480, and 20,480 samples, respectively. The proposed neural network is implemented on a system with NVIDIA GeForce GTX 2080Ti GPU. Training is conducted with a batch size of 512 over 50 epochs, using the Adam optimizer for parameter updates. Additionally, a learning rate schedule with a warm-up strategy is applied to improve convergence efficiency.
\vspace{-3mm}
\subsection{Performance Comparison} 

\begin{table}

\centering
\caption{the compared algorithms} 
\label{tab:compared algorithms}
\setlength{\tabcolsep}{1pt}
 \scalebox{0.6}{\begin{tabular}{|m{3.5cm}<{\centering}|m{1.8cm}<{\centering}|m{1.8cm}<{\centering}|m{1.9cm}<{\centering}|}
\hline
\textbf{Abbr.}& \textbf{Antenna selection method}& \textbf{Analog beamforming method}& \textbf{Digital beamforming method} \\ \hline
Proposed DyHBFNet&ASNet & ABFNet & DBFNet \\ \hline
FS(Vertical/Horizontal /Squared/Interlaced)\cite{Dynamic Subarrays} \& TransHBFNet & FS(Vertical /Horizontal /Squared /Interlaced)\cite{Dynamic Subarrays} & ABFNet & DBFNet \\ \hline
RS \& TransHBFNet & RS & ABFNet & DBFNet \\ \hline
Fully connnected \& TransHBFNet & Fully connnected  & ABFNet & DBFNet \\ \hline
ASNet \& BA \& DBFNet & ASNet & BA\cite{wmh} & DBFNet \\ \hline
ASNet \& ABFNet \& ZF & ASNet & ABFNet & ZF \\ \hline
ASNet \& ABFNet \& WMMSE & ASNet & ABFNet & WMMSE \\ \hline
ASNet \& ABFNet \& CNN-based WMMSE\cite{CNN-WMMSE} & ASNet & ABFNet & CNN-based WMMSE\cite{CNN-WMMSE}\\ \hline
ASNet \& HBF-Net\cite{Deep Unsupervised Learning} & ASNet & \multicolumn{2}{c|}{HBF-Net\cite{Deep Unsupervised Learning}  }\\ \hline
E-HBF-Net-1/2 \cite{EE-CNN} &  \multicolumn{3}{c|}{E-HBF-Net-1/2\cite{EE-CNN} }\\ \hline
\end{tabular}}
\end{table}

{To systematically evaluate the individual contributions of ASNet, ABFNet, and DBFNet to the proposed DyHBFNet, we conduct a series of ablation studies. For simplicity in subsequent discussions, ABFNet and DBFNet are collectively labeled as TransHBFNet. In Table \ref{tab:compared algorithms}, we employ the ASNet, one of the four fixed selection (FS) approaches in \cite{Dynamic Subarrays}, random selection (RS) approach, or fully connected approach as the antenna selection method. The ABFNet or beam alignment (BA)\cite{wmh} is used as the analog beamforming method. The digital beamforming methods include the DBFNet, Zero Forcing (ZF), WMMSE and CNN-based WMMSE\cite{CNN-WMMSE}.}\par
{Additionally, in Table \ref{tab:compared algorithms}, we compared TransHBFNet with HBF-Net\cite{Deep Unsupervised Learning} and antenna selection method with E-HBF-Net\cite{EE-CNN}. Specifically, to ensure fair comparison with E-HBF-Net\cite{EE-CNN}, we adopt two approaches: E-HBF-Net-1 implements the dynamic subarray architecture in \cite{EE-CNN}, which allows arbitrary connection between each antenna and RF chains, and is trained with the loss function defined in \cite{EE-CNN}. {E-HBF-Net-2 only adopts the CNN structure in \cite{EE-CNN},} while keeping the subarray architecture and loss function consistent with our letter.}  

\par As shown in Fig.~\ref{SE performance}\subref{fig_first_case}, the proposed ASNet architecture achieves significant performance gains compared to the conventional FS and RS approaches. {Fig.~\ref{SE performance}\subref{fig_second_case} shows that the ABFNet demonstrates superior performance over the BA algorithm and the DBFNet design surpasses ZF, traditional WMMSE algorithm, and CNN-based WMMSE\cite{CNN-WMMSE}. Also, the proposed TransHBFNet performs better SE than HBF-Net\cite{Deep Unsupervised Learning}.} Moreover, the proposed scheme demonstrates consistent performance advantages over baseline methods across varying user numbers ($K = 4$ and $K = 8$), highlighting the architecture's scalability.\par
\begin{figure}[!t]

\centering
\includegraphics[width=0.2\textwidth]{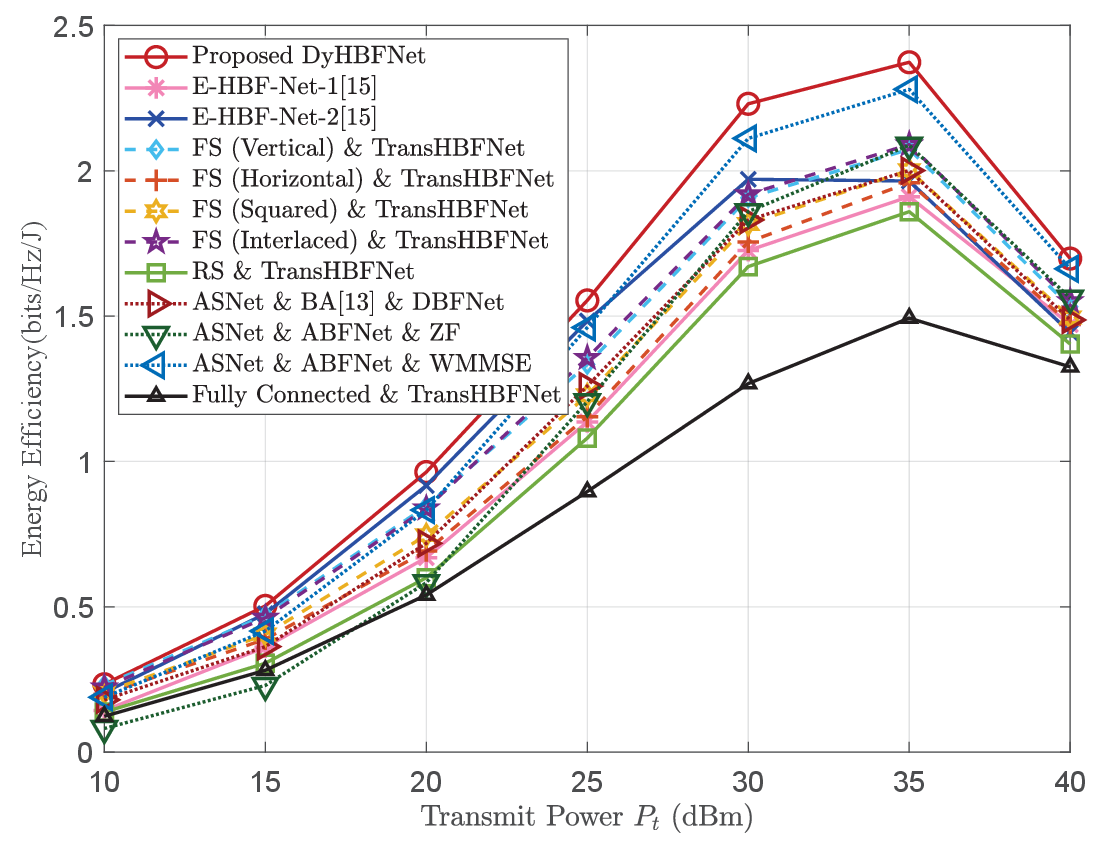}
\vspace{-3mm}
\caption{EE performance comparison versus $P_t$.}
\label{EE performance}
\vspace{-3mm}
\end{figure}

Furthermore, considering that both RF components and baseband processing contribute to the overall power consumption in MIMO systems, the EE can be expressed as  
$ \eta = \frac{R}{\frac{P_t}{\varepsilon} + P_{\text{BB}} + N_{\text{RF}}P_{\text{RF}} + N_t(P_{\text{PS}} + P_{\text{SW}})}$,
where $\varepsilon = 0.37$ represents the efficiency of the power amplifier, $P_{\text{BB}} = 1  \text{W}$ denotes the power consumption of baseband processing, and $P_{\text{RF}} = 300  \text{mW}$, $P_{\text{PS}}=40 \text{mW}$, and $P_{\text{SW}}= 5\text{mW}$ are the power consumption of each RFC, PS, and switch, respectively\cite{Energy Efficient}. {As shown in Fig.~\ref{EE performance}, all subarray strategies exhibit superior EE compared to the fully connected strategy. Among these, the proposed DyHBFNet achieves the highest EE, demonstrating its capability to reduce energy consumption effectively.}\par

\begin{figure}[!t]
\centering
\includegraphics[width=0.2\textwidth]{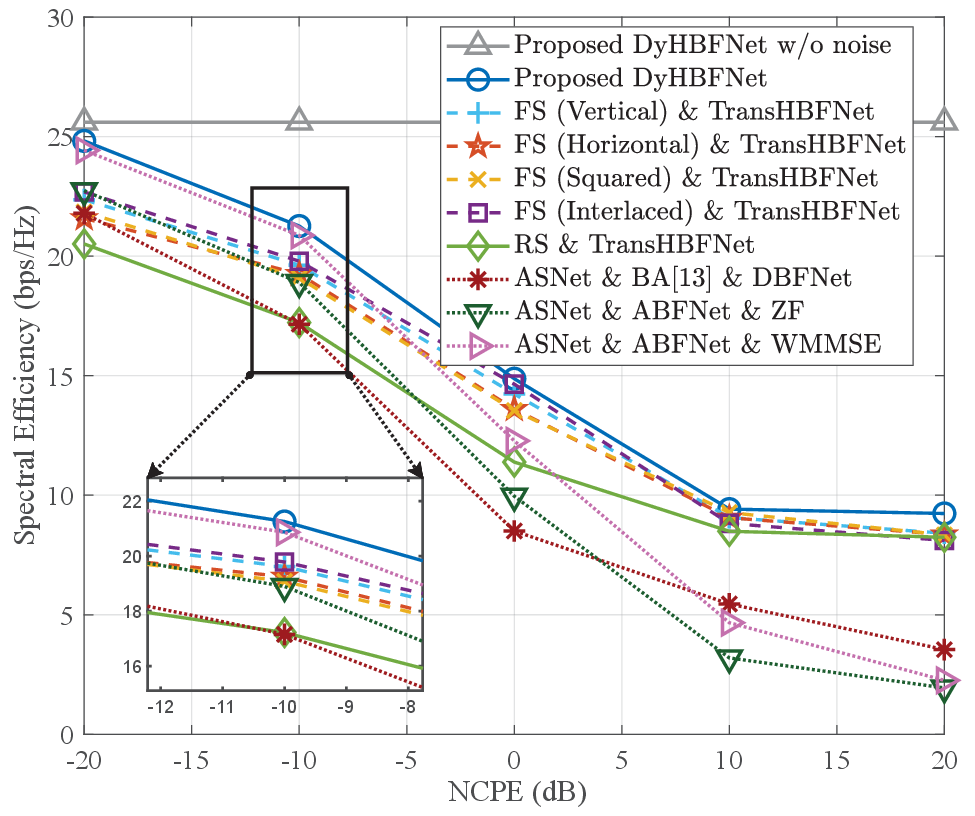}
\vspace{-4mm}
\caption{SE performance comparison under imperfect CSI scenario.}
\label{noise}
\vspace{-4mm}
\end{figure}
\begin{table}[!t]
    
    \belowrulesep=0pt
    \aboverulesep=0pt
    \setlength{\tabcolsep}{2pt}
    \centering
    \caption{Neural Network Complexity Comparison}
    \label{complexity comparision}
    \scalebox{0.6}{\begin{tabular}{c|c|c|m{3.2cm}<{\centering}}  
        \toprule
        & \textbf{Proposed DyHBFNet}& \textbf{E-HBF-Net-1/2\cite{EE-CNN}}& \textbf{ASNet \& ABFNet \& CNN-based WMMSE\cite{CNN-WMMSE}} \\ \midrule
        FLOPs & 133.89M & 101.98M & 43.48M \\ 
        Parameters & 18.26M & 70.00M &14.54M \\  
        Runtime & 11.82 ms & 3.56 ms & 9.56ms \\ \bottomrule  
    \end{tabular}}
\vspace{-5mm}
\end{table}

{On the other hand, the CSI in near-space communications is subject to unique perturbations induced by platform dynamics and upper atmospheric turbulence, and Fig.~\ref{noise} evaluates the robustness under such imperfect CSI conditions.} The normalized channel perturbation error (NCPE) is defined as $\text{NCPE} = \frac{\sum_{k=1}^K \sum_{m=1}^{N_c}\| \mathbf{h}[k,m] - \mathbf{h}_{\text{per}}[k,m] \|_2^2}{\sum_{k=1}^K \sum_{m=1}^{N_c} \| \mathbf{h}[k,m] \|_2^2}$, where the imperfect CSI matrix is modeled as $\mathbf{h}_{\text{per}}[k,m] = \mathbf{h}[k,m] + \mathbf{n}_{\text{per}}[k,m] $ and $\mathbf{n}_{\text{per}}[k,m] \sim \mathcal{CN}(0,\sigma^2_\text{per})$ is the perturbation noise \cite{syw}. The proposed scheme demonstrates remarkable error resilience, as its Transformer structure effectively exploits the correlations across subcarriers to extract robust channel features from contaminated inputs. \par
{Additionally, the network complexity of the proposed DyHBFNet and its counterparts is compared in Table \ref{complexity comparision}. In terms of FLOPs, the proposed DyHBFNet is higher than the other two algorithms, which can be attributed to its adoption of the more complex Transformer architecture.} For parameters, the proposed DyHBFNet is lower than E-HBF-Net-1/2 but slightly higher than ASNet \& ABFNet \& CNN-based WMMSE\cite{CNN-WMMSE}. This is because when extended to multi-carrier scenarios, E-HBF-Net-1/2 \cite{EE-CNN} significantly increases the parameter count of the FC layer at the final output stage to adapt to multi-carrier processing requirements. The proposed DyHBFNet has a longer runtime compared to the other algorithms, which is consistent with its higher computational complexity resulting from the Transformer architecture. Although the proposed DyHBFNet exhibits higher FLOPs and longer runtime compared to other algorithms, it achieves superior SE and EE.
\vspace{-2mm}
\section{Conclusion}
In this paper, we proposed a Transformer-based DyHBFNet for HAD architecture with dynamic subarray in near-space airship-borne communications. The proposed network facilitates the advantages of dynamic connections between antenna elements and RFCs by three dedicated DL modules for analog beamforming, antenna selection, and digital beamforming, respectively. Extensive simulations demonstrated that the proposed scheme achieved substantial improvements in both SE and EE compared to conventional methods, especially with strong robustness against imperfect CSI.

\vfill

\end{document}